\documentclass{eptcs}
\providecommand{\event}{PLACES 2015} 
\usepackage{breakurl}                

\usepackage{setspace}

\usepackage[labelfont=bf]{caption}
\usepackage{subcaption}
\usepackage{colortbl}%
  
\usepackage{wasysym}

\usepackage{graphicx}
\usepackage{xcolor}

\usepackage{listings}

\newcommand{\ie}{{\em i.e.,~}}
\newcommand{\eg}{{\em e.g.,~}}

\lstdefinelanguage{Scala}%
{morekeywords={abstract,case,catch,char,class,%
    def,else,extends,final,%
    if,import,%
    match,module,new,null,object,override,package,private,protected,%
    public,return,super,this,throw,trait,try,type,val,var,with,implicit,%
    macro,sealed,%
  },%
  sensitive,%
  morecomment=[l]//,%
  morecomment=[s]{/*}{*/},%
  morestring=[b]",%
  morestring=[b]',%
  showstringspaces=false%
}[keywords,comments,strings]%

\title{Distributed Programming via Safe Closure Passing}
\author{Philipp Haller
\institute{KTH Royal Institute of Technology\\ Sweden}
\email{phaller@kth.se}
\and
Heather Miller
\institute{EPFL\\
Switzerland}
\email{heather.miller@epfl.ch}
}
\def\titlerunning{Safe Closure Passing}
\def\authorrunning{P. Haller \& H. Miller}
\begin{document}
\maketitle

\begin{abstract}
Programming systems incorporating aspects of functional programming, e.g.,
higher-order functions, are becoming increasingly popular for large-scale
distributed programming. New frameworks such as Apache Spark leverage
functional techniques to provide high-level, declarative APIs for in-memory
data analytics, often outperforming traditional ``big data'' frameworks like
Hadoop MapReduce. However, widely-used programming models remain rather
ad-hoc; aspects such as implementation trade-offs, static typing, and semantics
are not yet well-understood. We present a new asynchronous programming model
that has at its core several principles facilitating functional processing of
distributed data. The emphasis of our model is on simplicity, performance,
and expressiveness. The primary means of
communication is by passing functions (closures) to distributed, immutable
data. To ensure safe and efficient distribution of closures, our model
leverages both syntactic and type-based restrictions. We report on a prototype
implementation in Scala. Finally, we present preliminary experimental results
evaluating the performance impact of a static, type-based optimization of serialization.
\end{abstract}

\section{Introduction}
\label{sec:introduction}

Programming systems for large-scala data processing are increasingly embracing
functional programming, \ie programming with first-class functions and
higher-order functions. Arguably, one of the first widely-used programming
models for ``big data'' processing making use of concepts from functional
programming is Google's MapReduce~\cite{DeanG08}. Indeed, \cite{Lammel08}
shows a precise executable semantics of MapReduce in Haskell. While leveraging
functional programming \emph{concepts}, popular implementations of the
MapReduce model, such as Hadoop MapReduce\footnote{See \url{http://hadoop.apache.org/}}
for Java, have been developed without making
use of functional \emph{language features} such as closures. In contrast, a new generation
of programming systems for large-scale data processing, such as
Apache Spark~\cite{Zaharia2012},
Twitter's Scalding,\footnote{See \url{https://github.com/twitter/scalding/}}
and Scoobi\footnote{See \url{http://nicta.github.io/scoobi/}} build on functional
language features in order to provide high-level, declarative APIs.

However, these programming systems suffer from several problems that
negatively affect their usage, maintenance, and optimization:

\begin{itemize}

\item Their APIs cannot statically prevent \emph{common usage errors}. As a result, users are often
      confronted with runtime errors that are hard to debug. A common example is
      unsafe closure serialization~\cite{MillerHO14}.

\item Typically, only high-level user-facing abstractions are statically
      typed. The absence of static types in lower layers of the system makes
      \emph{maintenance} tasks, such as code refactorings, more difficult.

\item The absence of certain kinds of static type information precludes systems-centric \emph{optimizations}.
      Importantly, type-based static meta-programming enables fast serialization~\cite{MillerHBO13}, but this is
      only possible if also lower layers (namely those dealing with object serialization) are statically typed.
      Several studies~\cite{Carpenter1999,Maassen1999,Philippsen2000,Welsh2000} report on the high
      overhead of serialization in widely-used runtime environments such as the JVM. This overhead is so
      important in practice that popular systems, like Spark~\cite{Zaharia2012} and Akka~\cite{Akka},
      leverage alternative serialization frameworks such as Protocol Buffers (Google),
      Apache Avro~\cite{Avro}, or Kryo~\cite{Kryo}.

\end{itemize}

\paragraph{Contributions}

This paper makes the following contributions:
\begin{itemize}

\item A new asynchronous programming model, called SCP (``safe closure passing''), for
      functional processing of distributed data (see Sec.~\ref{sec:model}). We propose to
      address the above problems through a novel combination of: (a) \emph{safe closures;}
      to prevent common usage errors. Closures that are not guaranteed to be serializable
      are rejected at compile time; (b) \emph{a statically-typed implementation of a generic
      distributed, persistent data structure.} Preserving static types through more system
      layers improves maintainability and enables type-based optimizations.

\item An implementation of the SCP model in Scala (see Sec.~\ref{sec:impl}). In addition,
      we report on preliminary experimental experience using SCP to evaluate the end-to-end
      performance impact of a type-based optimization of serialization.

\end{itemize}

An important goal of our model is to better understand programming systems such as Spark, Scalding, and
Scoobi, by incorporating several of their core principles; namely, immutable
distributed data and distributed closure passing. By focussing on simplicity,
expressiveness, and performance (and ignoring many of the more ad-hoc
refinements of the mentioned programming models) our
programming model--together with its prototype implementation--enables exploring
implementation trade-offs, and capturing the semantics of the core constructs more
precisely.

To ensure safe and efficient distribution of closures, our model leverages
both syntactic and type-based restrictions. For instance, closures sent to
remote nodes are required to conform to the restrictions imposed by the
so-called ``spore'' abstraction that the authors presented in previous
work~\cite{MillerHO14}. Among others, the syntax and static semantics of
spores can guarantee the absence of runtime serialization errors due to
closure environments that are not serializable.

The following Section~\ref{sec:spores} provides some required background.
Sections~\ref{sec:model} (programming model) and~\ref{sec:impl} (implementation)
present our main contributions. Section~\ref{sec:related} relates to prior work.
Section~\ref{sec:conclusion} summarizes future work and concludes.

\section{Background: Spores}
\label{sec:spores}

Spores~\cite{MillerHO14} are a closure-like abstraction and type system which aims to give users
a principled way of controlling the environment which a closure can capture.
This is achieved by (a) enforcing a specific syntactic shape which dictates
how the environment of a spore is declared, and (b) providing additional type-checking
to ensure that types being captured have certain properties. A crucial insight of
spores is that, by including type information of captured
variables in the type of a spore, type-based constraints for captured
variables can be composed and checked, making spores safer to use in a
concurrent, distributed, or in arbitrary settings where closures must be
controlled.





\subsection{Spore Syntax}
\label{sec:spore-syntax}

\setlength{\belowcaptionskip}{-6pt}
\begin{figure}[t!]
\centering
\includegraphics[width=7.4cm]{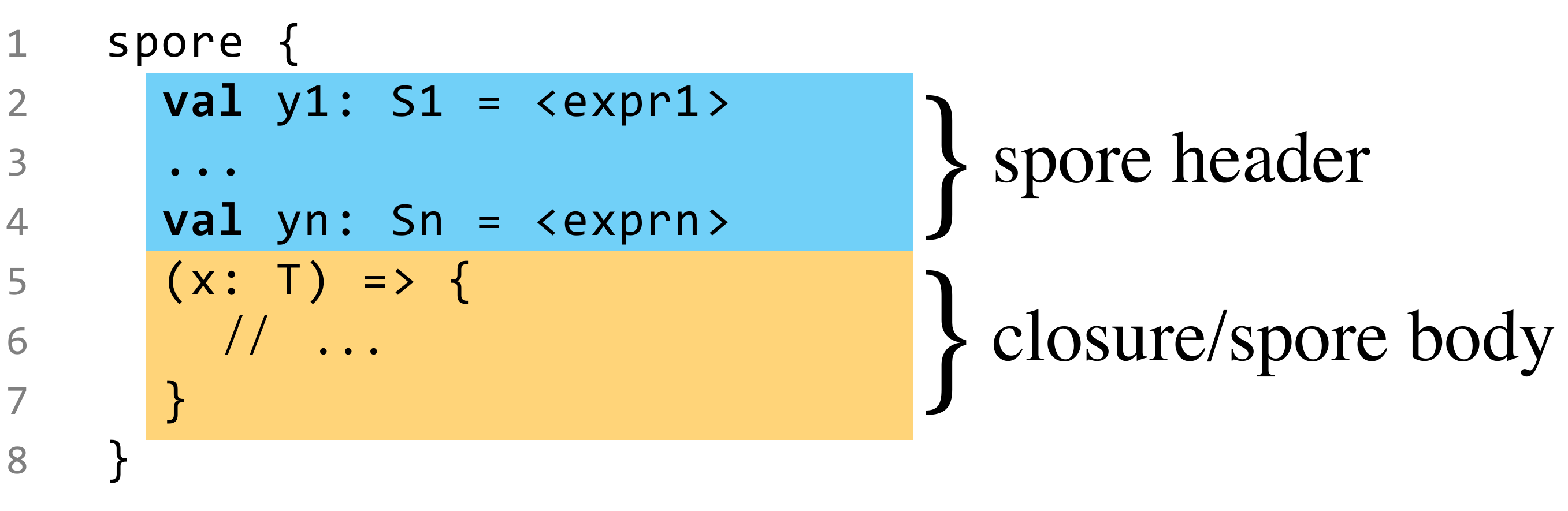}
\caption{The syntactic shape of a spore.}
\label{fig:spore-shape}
\end{figure}
\setlength{\belowcaptionskip}{0pt}

A spore is a closure with a specific shape that dictates how the environment
of a spore is declared. The shape of a spore is shown in Figure~\ref{fig:spore-shape}.
A spore consists of two parts:
\begin{itemize}
\item {\bf the spore header}, composed of a list of value definitions.
\item {\bf the spore body} (sometimes referred to as the ``spore closure''), a regular closure.
\end{itemize}


The characteristic property
of a spore is that the {\em spore body} is only allowed to access its
parameter, the values in the spore header, as well as top-level singleton objects
(public, global state). In particular, the spore closure is not allowed to
capture variables in the environment. Only an expression on the right-hand
side of a value definition in the spore header is allowed to capture
variables.

By enforcing this shape, the environment of a spore is always declared
explicitly in the spore header, which avoids accidentally capturing
problematic references. Moreover, importantly for object-oriented languages, it is no
longer possible to accidentally capture the \verb|this| reference.

\begin{figure}
\begin{subfigure}{.5\textwidth}
  \centering
  \begin{lstlisting}
  {
    val y1: S1 = <expr1>
    ...
    val yn: Sn = <exprn>
    (x: T) => {
      // ...
    }
  }
  \end{lstlisting}
  \caption{A closure block.}
  \label{fig:normal-block}
\end{subfigure}%
\begin{subfigure}{.5\textwidth}
  \centering
  \begin{lstlisting}
  spore {
    val y1: S1 = <expr1>
    ...
    val yn: Sn = <exprn>
    (x: T) => {
      // ...
    }
  }
  \end{lstlisting}
  \caption{A spore.}
  \label{fig:normal-spore-shape}
\end{subfigure}%
\vspace{1mm}
\caption{The evaluation semantics of a spore is equivalent to that of a closure, obtained by simply leaving out the \texttt{spore} marker.}
\label{fig:evaluation-semantics}
\end{figure}

\subsubsection{Evaluation Semantics}

The evaluation semantics of a spore is equivalent to a closure
obtained by leaving out the \verb|spore| marker, as shown in
Figure~\ref{fig:evaluation-semantics}. In Scala, the block shown in
Figure~\ref{fig:normal-block} first initializes
all value definitions in order and then evaluates to a closure that captures
the introduced local variables \verb|y1, ..., yn|. The corresponding spore,
shown in Figure~\ref{fig:normal-spore-shape} has the exact same evaluation
semantics. Interestingly, this closure shape is already used in production
systems such as Spark~\cite{Zaharia2012} in an effort to avoid problems with accidentally
captured references like \verb|this|. However, in systems like Spark, the
above shape is merely a convention that is not enforced.

\vspace{2mm}
\subsection{The \texttt{Spore} Type}
\label{sec:spore-type}


Figure~\ref{fig:spore-type} shows Scala's arity-1 function type and the arity-1 spore type.\footnote{For simplicity, we omit \texttt{Function1}'s definitions of the \texttt{andThen} and \texttt{compose} methods.}
Functions are contravariant in their argument type \verb|A| (indicated using
\verb|-|) and covariant in their result type \verb|B| (indicated
using \verb|+|). The \verb|apply| method of \verb|Function1| is abstract; a concrete implementation applies
the body of the function that is being defined to the parameter \verb|x|.

Individual spores have {\em refinement types} of the base \verb|Spore| type, which, to be compatible with normal Scala functions,
is itself a subtype of \verb|Function1|. Like functions, spores are contravariant in their argument type \verb|A|, and
covariant in their result type \verb|B|. Unlike a normal function,
however, the \verb|Spore| type additionally contains information about
\textit{captured} and \textit{excluded} types. This information is represented
as (potentially abstract) \verb|Captured| and \verb|Excluded| type members. In a
concrete spore, the \verb|Captured| type is defined to be a tuple with the types of all captured variables.
\cite{MillerHO14}~discusses the \verb|Excluded| type member in detail.

\begin{figure}[t!]
\begin{subfigure}{.5\textwidth}
  \centering
  \begin{lstlisting}
    trait Function1[-A, +B] {
      def apply(x: A):  B
    }
  \end{lstlisting}
  \caption{Scala's arity-1 function type.}
  \label{fig:function-arity1}
\end{subfigure}%
\begin{subfigure}{.5\textwidth}
  \centering
  \begin{lstlisting}
    trait Spore[-A, +B]
    extends Function1[A, B] {
      type Captured
      type Excluded
    }
  \end{lstlisting}
  \caption{The arity-1 \texttt{Spore} type.}
  \label{fig:spore-arity1}
\end{subfigure}%
\vspace{1mm}
\caption{The \texttt{Spore} type.}
\label{fig:spore-type}
\vspace{-2mm}
\end{figure}




\vspace{2mm}
\subsection{Basic Usage}
\label{sec:basic-usage}
\vspace{1mm}


\subsubsection{Definition}

A spore can be defined as shown in Figure~\ref{fig:captured-spore}, with its
corresponding type shown in Figure~\ref{fig:captured-type}. As can be seen,
the types of the environment listed in the spore header are
represented by the \verb|Captured| type member in the spore's type.

\begin{figure}[t!]
\begin{subfigure}{.5\textwidth}
  \centering
  \begin{lstlisting}
    val s = spore {
      val y1: String = expr1;
      val y2: Int = expr2;
      (x: Int) => y1 + y2 + x
    }
  \end{lstlisting}
  \caption{A spore \texttt{s} which captures a \texttt{String} and an \texttt{Int} in its spore header.}
  \label{fig:captured-spore}
\end{subfigure}%
\begin{subfigure}{.5\textwidth}
  \centering
  \begin{lstlisting}
    Spore[Int, String] {
      type Captured = (String, Int)
    }
  \end{lstlisting}
  \caption{\texttt{s}'s corresponding type.}
  \label{fig:captured-type}
\end{subfigure}%
\vspace{1mm}
\caption{An example of the \texttt{Captured} type member. \\\textit{Note: we omit the
\texttt{Excluded} type member for simplicity; we discuss it in detail in~\cite{MillerHO14}.}}
\label{fig:captured-ex}
\vspace{-5mm}
\end{figure}

\subsubsection{Using Spores in APIs}

Consider the following method definition:

\begin{lstlisting}[numbers=none]
    def sendOverWire(s: Spore[Int, Int]): Unit = ...
\end{lstlisting}
\noindent
In this example, the \verb|Captured| (and \verb|Excluded|) type
member is not specified, meaning it is left abstract. In this case, so long as
the spore's parameter and result types match, a spore type is always
compatible, regardless of which types are captured.

Using spores in this way enables libraries to enforce the use
of spores instead of plain closures, thereby reducing the risk for common
programming errors, even in this very simple form.

\subsubsection{Composition}

Like normal functions, spores can be composed. By representing the environment
of spores using refinement types, it is possible to preserve the captured type
information of spores when they are composed.

For example, assume we are given two spores \verb|s1| and \verb|s2| with types:

\begin{lstlisting}[numbers=none]
    s1: Spore[Int, String] { type Captured = (String, Int) }
    s2: Spore[String, Int] { type Captured = Nothing }
\end{lstlisting}

\noindent The fact that the \verb|Captured| type in \verb|s2| is defined to be
\verb|Nothing| means that the spore does not capture anything (\verb|Nothing|
is Scala's bottom type). The composition of \verb|s1| and \verb|s2|, written
\verb|s1 compose s2|, would therefore have the following refinement type:

\begin{lstlisting}[numbers=none]
    Spore[String, String] { type Captured = (String, Int) }
\end{lstlisting}

\noindent Note that the \verb|Captured| type member of the result spore is equal to the
\verb|Captured| type of \verb|s1|, since it is guaranteed that the result
spore does not capture more than what \verb|s1| already captures. Thus, not
only are spores composable, but so are their (refinement) types.

\section{Programming Model}
\label{sec:model}

The programming model has a few basic abstractions at its center: first, the
so-called \emph{silo}. A silo is a typed data container. It is stationary in the
sense that it does not move between machines. A silo remains on the machine
where it was created. Data stored in a silo is typically loaded from stable
storage, such as a distributed file system. A program operating on data stored
in a silo can only do so using a reference to the silo, a so-called \emph{SiloRef}.
Similar to a proxy object, a SiloRef represents, and allows interacting with,
a silo possibly located on a remote node. Some programming patterns require
combining data contained in silos located on different nodes (\eg joins). To
support such patterns, our model includes a \emph{pumpTo} primitive for emitting
data to silos on arbitrary nodes (explained further below).

A SiloRef has the following main operations:
\begin{verbatim}
trait SiloRef[T] {
  def apply(s: Spore[T, S]): SiloRef[S]
  def send(): Future[T]
}
\end{verbatim}

\paragraph{Apply}

The \verb|apply| method takes a spore, a kind of closure (see Section~\ref{sec:spores} for an overview),
that is to be applied to the data in the silo of the receiver SiloRef. Rather than immediately sending
the spore across the network, and waiting for the operation to finish, the
apply method is \emph{lazy}. It immediately returns a SiloRef that refers to the
result silo.

To realize something like the \verb|map| combinator of a (distributed) collection using \verb|apply|,
it is helpful to think of the spore argument to \verb|apply| (``s'') as the composition of a
user-defined function passed to the \verb|map| combinator with the actual implementation of \verb|map|:

\begin{verbatim}
val ref: SiloRef[List[Int]] = ...
val userFun: Int => String = ...
val mapFun: (Int => String) => List[Int] => List[String] = ...
val ref2: SiloRef[List[String]] = ref.apply(mapFun(userFun))
\end{verbatim}

In the above example, the higher-order \verb|mapFun| function is expressed in curried
style where the user's function argument is passed as the first argument.
Applying \verb|mapFun| to a function of type \verb|Int => String| returns a function of type
\verb|List[Int] => List[String]|.

The result of invoking \verb|apply| is another SiloRef, which has a reference to the
spore and the SiloRef that it was derived from. Note that this is semantically
the same as programming with normal functional data structures, where a new
data structure is defined by a transformation of an original data structure.

\paragraph{Send}

The \verb|send| method takes no argument, and returns a future. Unlike
\verb|apply|, \verb|send| is \emph{eager} (readers familiar with the concept
of \emph{views} might recognize a similarity to forcing a view). That is, it
sends whatever operations are queued up (by invocations of \verb|apply|) on a
given SiloRef to the node that contains the corresponding silo, and kicks off
the materialization of the result silo. Once the materialization is done, the
future returned by \verb|send| is completed. Example:

\begin{verbatim}
val ref2 = ref1.apply(s)  // lazy
val fut  = ref2.send()    // eager
\end{verbatim}
\noindent
The invocation of \verb|send| kicks off the following sequence of actions:
\begin{enumerate}
\item A ``send'' control message is sent to the node where \verb|ref2|'s silo is located.
\item Since \verb|ref2| is derived from \verb|ref1|, \verb|ref1|'s silo is located on the same node. Thus, the runtime demands
      \verb|ref1| to be materialized; once this is done, spore \verb|s| is applied, populating \verb|ref2|'s silo (on the same node).
\item Once \verb|ref2|'s silo is materialized, its data is sent to the node executing the \verb|send|, completing \verb|fut|.
\end{enumerate}
\noindent
Note that since a \verb|send| operation sends the data of a silo across the network, it should only be invoked on silos containing small bits of data.

\paragraph{pumpTo}

The SiloRef singleton object provides an additional method for combining silos
storing collections:

\begin{verbatim}
def pumpTo[T <: Traversable[U], V, R](
      p: Place,
      silo1: SiloRef[T],
      silo2: SiloRef[T],
      fun: Spore[(U, Emitter[V]), Unit],
      bf: BuilderFactory[V, R]): SiloRef[R]
\end{verbatim}
\noindent
The \verb|pumpTo| method requires the silos \verb|silo1| and \verb|silo2| to contain collections of
element type \verb|U| \newline (\verb|Traversable[U]|). Using \verb|pumpTo|, the elements (of type \verb|U|) of the
two silos are passed one-by-one to the user-provided spore \verb|fun|. This spore
takes a pair as argument containing two components: first, a single element of
one of the silo's collections, and second, an \emph{emitter} to which the spore can
output values of type \verb|V|. By emitting such elements, a new silo at the
destination (\verb|Place p|) is filled, yielding a collection of type \verb|R|. A
\verb|BuilderFactory[V, R]| provides the functionality for building a collection of
type \verb|R| based on elements of type \verb|V|. Finally, \verb|pumpTo| returns the SiloRef of the
silo that was created at \verb|Place p|.

Although conceptually related, the \verb|Emitter| and the \verb|BuilderFactory| address two
separate issues: the \verb|Emitter| provides a way to output elements from the source
silo; the \verb|BuilderFactory| provides a way to input data into a newly created
silo at the destination.

An \verb|Emitter| is a simple trait which allows the spore parameter \verb|fun| of \verb|pumpTo| to
emit zero, one, or multiple values per element of a silo's collection, using an
\verb|emit| function:
\begin{verbatim}
def emit(v: T)(implicit p: Pickler[T]): Unit
\end{verbatim}
\noindent
Note that \verb|Emitter| differs from the well-known \emph{observable} abstraction~\cite{Meijer12}
in one important way: the emit method requires an implicit type-specific
\emph{pickler}. In Scala, type-specialized picklers enable fast serialization
through compile-time meta-programming~\cite{MillerHBO13}. Thus, \verb|Emitter| is an
abstraction specially designed for distributed programming. The main reason
why the pickler is required already at this point is that we would like to enable
picklers to be \emph{specialized} to the type of the pickled values. However,
this means a pickler has to be constructed at the point when the static type
of the emitted value is still available (essentially, before the
value loses its type when treated as generic data to be sent across the
network).

\subsection{Combining Multiple Silos}

Operations on distributed collections such as \emph{union}, \emph{groupByKey}, or \emph{join},
involve multiple data sets, possibly located on different nodes. In the
following we explain how such operations can be expressed using the introduced
primitives.

\paragraph{union}

The union of two unordered collections stored in two different silos can be
expressed directly using the above \verb|pumpTo| primitive.

\paragraph{join}

Suppose we are given two silos with the following types:
\begin{verbatim}
val silo1: SiloRef[List[A]]
val silo2: SiloRef[List[B]]
\end{verbatim}
\noindent
as well as two hash functions computing hashes for elements of type \verb|A| and \verb|B|, respectively:
\begin{verbatim}
val hashA: A => K = ...
val hashB: B => K = ...
\end{verbatim}
\noindent
The goal is to compute the hash-join of \verb|silo1| and \verb|silo2|:
\begin{verbatim}
val hashJoin: SiloRef[List[(K, (A, B))]] = ???
\end{verbatim}
\noindent
To be able to use \verb|pumpTo|, the types of the two silos first have to be made equal, through initial \verb|apply| invocations:
\begin{verbatim}
val silo12: SiloRef[List[(K, Option[A], Option[B])] =
      silo1.apply { x => (hashA(x), Some(x), None) }
val silo22: SiloRef[List[(K, Option[A], Option[B])] =
      silo2.apply { x => (hashB(x), None, Some(x)) }
\end{verbatim}
\noindent
Then, we can use \verb|pumpTo| to create a new silo (at some destination place), which contains the elements of both \verb|silo12| and \verb|silo22|:
\begin{verbatim}
val combined = SiloRef.pumpTo(destPlace, silo12, silo22,
                              (elem, emitter) => emitter.emit(elem),
                              listBuilderFactory[...])
\end{verbatim}
\noindent
The combined silo contains triples of type \verb|(K, Option[A], Option[B])|. Using an
additional \verb|apply|, the collection can be sorted by key, and adjacent triples be
combined, yielding finally a \newline \verb|SiloRef[List[(K, (A, B))]]| as required.

\paragraph{Partitioning and groupByKey}

A \emph{groupByKey} operation on a group of silos containing collections needs to create multiple result silos, on each node, with ranges of keys supposed to be shipped to destination nodes. These destination nodes are determined using a partitioning function. Our goal, concretely:
\begin{verbatim}
val groupedSilos = groupByKey(silos)
\end{verbatim}
\noindent
Furthermore, we assume that $\texttt{silos.size} = N$ where $N$ is the number of nodes,
with nodes $N_1$, $N_2$, etc. We assume each silo contains an unordered collection
of key-value pairs (a multi-map). Then, \emph{groupByKey} can be implemented as
follows:
\begin{itemize}
\item For each node $N_i$, the master node creates $N$ SiloRefs.
\item Each node $N_i$ applies a \emph{partitioning function} (example: \verb|hash(key) mod N|) to
      the key-value pairs in its silo, yielding $N$ (local) silos.
\item Using \verb|pumpTo|, each pair of silos containing keys of the same range can be combined
      and materialized on the right destination node.
\end{itemize}

\section{Implementation and Preliminary Experimental Results}
\label{sec:impl}

We have developed a prototype\footnote{See \url{https://github.com/heathermiller/f-p}}
of the SCP model in Scala, which builds on our earlier work on
Scala Pickling~\cite{MillerHBO13} and Spores~\cite{MillerHO14}.
The implementation does not require extensions to the Scala language or compiler; it
is developed using the current stable Scala release 2.11.

We have used our implementation to measure the impact of
compile-time-generated serializers~\cite{MillerHBO13} on end-to-end application
performance. We ran our experiments on a 2.3 GHz Intel Core i7
with 16 GB RAM under Mac OS X 10.9.5 using Java HotSpot Server 1.8.0-b132.
In our benchmark application, a group of 4 silos is distributed
across 4 different nodes/JVMs. The silos are first transformed using
\emph{map}, and then using \emph{groupBy}. For an input size of 100'000 ``person''
records, the use of compile-time-generated serializers resulted in
an overall speedup of about 48\% with respect to the same system but without using
compile-time-generated serializers.

\section{Related Work}
\label{sec:related}

Cloud Haskell~\cite{CloudHaskell} leverages guaranteed-serializable, static closures
for a message-passing communication model inspired by Erlang. In contrast, in
our model spores are sent between passive, persistent silos. Closures and
continuations in Termite Scheme~\cite{Termite} are always serializable;
references to non-serializable objects (like open files) are automatically
wrapped in processes that are serialized as their process ID. Similar to Cloud
Haskell, Termite is inspired by Erlang. In contrast to Termite, SCP is statically typed,
enabling advanced type-based optimizations. In non-process-oriented models,
parallel closures~\cite{ParallelClosures} and RiverTrail~\cite{HerhutHSS13} address
important safety issues. SCP integrates a distributed, persistent data structure.
Other prior work related to spores is discussed in~\cite{MillerHO14}.

\section{Conclusion and Future Work}
\label{sec:conclusion}

We have presented a new asynchronous distributed programming model.
A novel combination of (a) closures with syntactic and semantic restrictions and
(b) abstractions for distributed data ``silos'' prevents usage errors common in
widely-used ``big data'' frameworks.
We have implemented our model in Scala; preliminary experimental
results evaluate the performance impact of a static type-based optimization. In
future work, we intend to expand the practical
experiments and explore the impact of implementation trade-offs. Furthermore, we plan
to exploit the simplicity of the programming model for a formal treatment of its
properties.

\nocite{*}
\bibliographystyle{eptcs}
\bibliography{bib}
\end{document}